boilerplate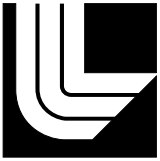

LLNL-JRNL-665381LAWRENCE LIVERMORE NATIONAL LABORATORY# Scaling the Yield of Laser-Driven Electron-Positron Jets to Laboratory Astrophysical Applications

H. Chen, F. Fiuza

December 11, 2014

Physical Review Letters

**Scaling the Yield of Laser-Driven Electron-Positron Jets to Laboratory Astrophysical Applications**

Hui Chen[1], F. Fiuza[1], A. Link[1], A. Hazi[1], M. Hill[2], D. Hoarty[2], S. James[2], S. Kerr[3], D. D. Meyerhofer[4], J. Myatt[4], J. Park[1], Y. Sentoku[5], G. J. Williams[1]

1. Lawrence Livermore National Laboratory, CA 94550, USA

2. Directorate of Science and Technology, AWE plc, Reading, RG7 4PR, UK

3. University of Alberta, Alberta T6G 2R3, Canada

4. Laboratory for Laser Energetics, University of Rochester, Rochester, NY 14623, USA

5. University of Nevada, Reno, Nevada 89557, USA

**Abstract:**

**We report new experimental results obtained on three different laser facilities that show directed laser-driven relativistic electron-positron jets with up to 30 times larger yields than previously obtained and a quadratic (~ $E_L^2$) dependence of the positron yield on the laser energy. This favorable scaling stems from a combination of higher energy electrons due to increased laser intensity and the recirculation of MeV electrons in the mm-thick target. Based on this scaling, first principles simulations predict the possibility of using such electron-positron jets, produced at upcoming high-energy laser facilities, to probe the physics of relativistic collisionless shocks in the laboratory.**

Relativistic electron-positron pair plasmas are ubiquitous in high-energy astrophysical environments, such as Gamma Ray Bursts (GRBs), Active Galactic Nuclei and Pulsar Wind Nebulae. These systems share the common observational feature of broad nonthermal spectrum radiation, which is usually assumed to be produced by energetic particles accelerated at relativistic shocks [1] or during magnetic reconnection [2].

For example, in the fireball model of GRBs, a massive black hole or neutron star produces a relativistic fireball in the form of an electron-positron plasma and radiation. The gamma-rays are produced by synchrotron or inverse Compton from high-energy, Fermi-accelerated electrons in optically thin relativistic shocks within this fireball [3]. Plasma processes mediate the interaction of relativistic, weakly magnetized pair shells in GRBs. In particular, the Weibel, or current filamentation, instability [4] is believed to play a dominant role in the initial generation of near-equipartition magnetic fields and formation of shocks [5]. The growth rate of the instability for two counter-streaming symmetric pair flows is given by $\Gamma = \sqrt{2}\frac{v_0}{c}\omega_p$, where $\omega_p = \sqrt{4\pi n_0 e^2/(\gamma_0 m_e)}$ is the relativistic plasma frequency, $v_0$ and $\gamma_0$ are the flow velocity and Lorentz factor, $c$ is the speed of light, $n_0$ is the plasma density, $m_e$ is the electron mass, and $e$ is the elementary charge. Studies using fully-kinetic Particle-In-Cell (PIC) simulations have shown that this instability can generate small-scale magnetic fields that deflect the incoming flows [6], thereby mediating the formation of a collisionless shock [7], the Fermi acceleration of particles [8], and the emission of synchrotron radiation [9]. These processes are critical to understand the radiation emission in GRBs and could be directly probed in the laboratory by colliding two counter-streaming electron-positron flows.

The possibility of probing the physics of relativistic shocks in the laboratory has motivated the pursuit of the generation of high-flux relativistic pair jets. The use of high-power lasers to produce jets of megaelectron-volt (MeV) electron-positron plasma has been demonstrated experimentally in recent years using solid, high-Z targets [10–12]. The generation of relativistic electron-positron pairs occurs through the Bethe-Heitler (B-H) process [13] and involves three steps: (i) the laser energy is transferred to relativistic electrons in the preformed plasma at the front of the solid target; (ii) these electrons convert part of their energy to MeV bremsstrahlung photons as they go through the target; (iii) the bremsstrahlung photons produce pairs in the field of the high-Z target nuclei. Positrons are then further accelerated by the sheath electric field at the rear side of the target [14], leading to the emission of a relativistic electron-positron jet. The characteristics of these jets make them well suited for the study of scaled astrophysical phenomena in laboratory experiments. In particular, (i) the energy of each electron-positron pair can reach from a few to tens of MeV, a range similar to that predicted for GRB fireballs, (ii) the emission is directed with a typical half-angle divergence [15] of 15°-20°, and (iii) the electron-positron jets have relatively high density due to the small plasma volumes (~mm$^3$) arising from the combination of small laser-target interaction size and short pulse duration (1 – 10 ps) [10]. However, the maximum yield obtained in previous experiments is still a few orders of magnitude below what would be needed for scaled laboratory astrophysics experiments, as dictated by the spatial scale and growth rate of the instability.

To use such pair plasma jets to study the development of the Weibel instability and the magnetic field dynamics associated with relativistic collisionless shocks, it is necessary to guarantee that the duration of the flows is greater than the typical time for instability growth, $\tau_0 > 1/\Gamma$, and that the transverse scale of the plasma is larger than the spatial scale of the instability, $R_0 > c/\omega_p$, where $R_0$ is the radius of the electron-positron flow. These requirements impose a lower limit on the electron-positron yield

$$N[10^{11}] > \begin{cases} 1.5(R_0[\text{mm}])^2\gamma_0/\tau_0[\text{ps}], & R_0[\text{mm}] \geq 0.42\,\tau_0[\text{ps}] \\ 0.27\gamma_0\tau_0[\text{ps}], & R_0[\text{mm}] < 0.42\,\tau_0[\text{ps}] \end{cases} \quad (1)$$

For a 10 ps long, mm-scale flow, the required pair yield is $> 10^{11}$-$10^{12}$.

To evaluate whether laser-produced pair plasma flows can meet the above requirement, we performed experiments at three high-energy laser facilities, Titan [16], Orion [17] and Omega EP [18], using laser energies ranging from 100 J to 1.5 kJ and pulse lengths of 1 ps and 10 ps. The lasers (with wavelength of 1.054 µm, $10^{6-7}$ intensity contrast ratio and focused using *f*/2 or *f*/3 off-axis-parabola) were incident on 1 mm thick, 2 mm diameter gold disc targets in all shots at angles less than 18° from the target normal. With the FWHM focal spot size of 10-30 µm, the laser intensities were between $3\times10^{18}$ – $1\times10^{20}$ W/cm$^2$. At the given laser contrast, the nanoseconds long prepulse at $10^{12\text{-}14}$ W/cm$^2$ intensity creates a preformed plasma with density scale-length of about 20 – 50 microns on the target surface with which the main pulse interacts and accelerates the electrons.

The energy and angular distribution of electrons, positrons and protons emitted from the target were measured with magnetic spectrometers [10, 11] placed at 0° - 90° angles from the target normal. The measured yield of relativistic pairs, *N*, as a function of the laser energy, $E_L$, as shown in Figure 1. The data points fall into two groups, for 1 and 10 ps. The best fit to the data (shown by the lines) in each group for the yield *N* as a function of laser energy $E_L$ gives: N=4(±9)×10$^5$ $E_L^{(2.3±0.4)}$ and N=2(±3)×10$^5$ $E_L^{(2.0±0.3)}$ for 1 and 10 ps, respectively. Both data groups show a non-linear, approximately quadratic ($N \sim E_L^2$) dependence of the positron yield on the incident laser energy. The absolute yield is higher for the 1 ps case, since it corresponds to a higher laser intensity for fixed laser energy and spot size. The uncertainty in the measured positron yield comes from the angular distribution (~20%), spectrometer calibration (~20%), image plate read-out (~15%), and the total error is about 30% for each data point. The highest yield, $N = 6 \times 10^{11}$, achieved on Omega EP using a 1.45 kJ, 10 ps laser pulse, is about 30 times higher than that obtained previously [11].

In order to understand the scaling obtained experimentally we compared the data with positron yields calculated by Monte-Carlo code GEANT4 [19], semi-analytical model described by Myatt et al. in [20], and PIC simulations using the code LSP [21]. All these models capture pair-production through B-H process, and electron slowing down by both collisions and bremsstrahlung emission.

GEANT4 simulations calculate the positron yield for a given electron distribution in the target. The electron distribution depends on the physics of laser absorption. Different electron acceleration mechanisms operate at in different regions of the plasma [22]. Near the critical density (defined as the density above which the laser does not propagate) acceleration occurs predominantly through the **JxB** mechanism, leading to the so-called ponderomotive scaling [23] for $T_e$ as a function of laser intensity, $I$ (in Wcm$^{-2}$): $T_e \approx 0.511 \times \left( \sqrt{1 + I\lambda^2/1.4 \times 10^{18}} - 1 \right)$ (MeV), where $\lambda$ is the laser wavelength in μm. In the under-dense region of the plasma, produced by target ablation from the laser pre-pulse, acceleration occurs due to stochastic processes [24] and $T_e$ is given by [25] $T_e \approx 1.5 \times \left(\sqrt{I\lambda^2/10^{18}}\right)$ (MeV) - hereafter referred to as Pukhov scaling. These two mechanisms represent the limiting cases of the laser-plasma interaction conditions present in our experiments and were used as inputs to GEANT 4 calculations and the semi-analytical model. The latter mechanism is responsible for the generation of the most energetic particles and therefore is expected to play the dominant role in pair production. Recent simulations for similar laser and plasma conditions confirm that indeed the most energetic particles originate within the underdense plasma region [26].

Fig. 2 shows a direct comparison between the experimental results and the numerical/analytical calculations in terms of the positron yield per kJ of laser energy as a function of laser intensity. The yield increases with laser intensity up to $2 \times 10^{20}$ W/cm$^{-2}$ for the 1 mm thick gold target used in this study; beyond this laser intensity, the increase rate may slow down as electrons become too energetic to effectively create positrons using this target. The intensity is calculated using the measured laser energy, the laser focal area containing 68% of the energy and the laser pulse length for each shot, with uncertainty (30%-50%) from laser energy (~5%-10%), pulse length (~10%-20%), and focal spot size (20%-40%). The calculated yields clearly fit the data better using the Pukhov $T_e$ scaling than the ponderomotive scaling. The fact that the most energetic electrons produce the positrons indicates the important role of electron acceleration from under-dense plasma in these experiments and is consistent with the fact that all three lasers produced appreciable sub-critical density pre-plasma due to their moderate contrast. At laser intensities greater than $10^{19}$ Wcm$^{-2}$, the GEANT4 calculations start to deviate from the experimental data. We should note that GEANT4 does not include generation of electromagnetic fields, which can cause electron refluxing in the target [27].

In order to take into account the role of electron refluxing, we used semi-analytical model [20] which includes the same pair-production physics of GEANT4 but includes an additional electron refluxing parameter $\eta$ that enables a fraction (0<$\eta$<1) of the laser-produced electrons to recirculate through the target (see Eqs. (8), (9), and (22) of Ref. [20]), leading to an increased population participating in the pair production. These refluxing electrons are not expected to interact with the laser, as has been observed with relatively thin targets [28, 29], because of the large (mm) target size, divergence of the electron beam [11] and small focal size of the laser.

The semi-analytical model agrees well with the data for high refluxing levels ($\eta$ = 1) suggesting that electron refluxing plays an important role in positron generation by effectively increasing the interaction of fast electrons in the target. This important effect, seen in thin targets [27], had not been realized before for mm thick targets at high intensities. Note that as the laser energy at fixed intensity decreases, positron production would move from the analytical model with $\eta$=1 to the GEANT4 calculation as less and less electrons reflux through the target.

To confirm the role of electron refluxing in the laser-target interaction, we performed PIC simulations using the LSP code [21] that self-consistently model the laser-plasma interaction, electron transport and pair production. LSP simulations thus provide a direct benchmark to the above-mentioned calculations. The laser-plasma interaction was simulated in 2D for a $8 \times 10^{19}$ Wcm$^{-2}$ laser-target interaction for a target with a pre-plasma obtained for the actual laser (Titan) contrast using hydrodynamics simulations. The laser plasma simulation was resolved with 20 cells per laser wavelength and 66 time-steps per laser period using temporal and focal profiles taken from on shot measurements with a preplasma profile which along the laser axis that can be parameterized by a 2.2 $\mu$m scale-length from solid density to critical density and 25 $\mu$m scale-length from critical density to 1/10 critical density. Multi-dimensional effects associated with self-focusing and channeling in the laser-plasma interaction results in very hot electron tail with temperature about 3 times ponderomotive scaling – similar to that of Pukhov scaling. The resulting hot electron source was fed to the 2D transport simulation of a full experimental scale target - a 2 mm diameter, 1 mm thick solid Au target embedded in a vacuum box spanning 5 mm in radius and 1.5 cm in length. Bremsstrahlung and B-H pair production were calculated at every time step, and self consistently evolved with the rest of the simulation. The transport simulation is used to investigate the role of electron recirculation, which naturally arises from the highest energy electrons leaving and charging the target, resulting in about 4% of total electrons escaping the target while the rest participate in the recirculation. This confirms the high level of refluxing present under these experimental conditions. To illustrate the role of the electron refluxing on the total yield we have repeated the same simulation turning off the electro-magnetic field solver, which therefore removes the electrostatic sheath field causing the electron refluxing. The results for these two cases are compared to the experimental data in Fig. 2 and support the refluxing hypothesis.

Having identified the dominant physics associated with the scaling of positron yield with laser and plasma parameters, we now turn to its implication for using laser-produced pair plasma flow to study the shock physics relevant to GRBs. The favorable scaling of the positron yield with laser energy obtained in our experiments suggests that, at a laser intensity and pulse duration comparable to what is currently available at Omega EP, near-future 10 kJ class lasers would provide 100 times higher positron yield (up to $N_{e+} \sim 10^{14}$) than the present record ($N_{e+} \sim 10^{12}$). Lasers with such energies are being built, for example, the ARC laser on NIF [30] and the LFEX laser on GEKKO [31].

Laboratory experiments at such laser facilities can be used to study the generation of internal shocks and their role in magnetic field amplification, particle acceleration, and radiation emission. The microphysics of this collisionless interaction can be directly scaled between laboratory conditions and astrophysical scenarios [32]. Strong magnetic fields grow from the thermal fluctuation level due to the Weibel instability and scatter the particles of the incoming particles flows leading to shock formation. The shock formation time for relativistic pair plasmas [33, 34] can be estimated as

$$\tau_{sh}[\text{ps}] \simeq 12 \sqrt{\frac{\gamma_0}{n_0[10^{15}\text{cm}^{-3}]}} \left(1 + 4.9 \times 10^{-2} \log\left(\frac{1}{T_e[\text{MeV}]}\sqrt{\frac{\gamma_0}{n_0[10^{15}\text{cm}^{-3}]}}\right)\right).$$

Based on the scaling of the number of pairs with laser energy obtained experimentally, and confirmed theoretically, for 10 ps lasers, we can estimate that the laser energy required to study the formation of a shock is

$$E_{L,shock}[\text{kJ}] \simeq 28.5\, R_0[\text{mm}] \sqrt{\frac{\gamma_0}{\tau_0[\text{ps}]}}. \tag{2}$$

This shows that 10 kJ class lasers, soon in operation, can be used to study the physics of relativistic electron-positron shocks for the first time in the laboratory. For example, NIF ARC is designed to have 12 kJ and a pulse duration ranging from 1-30 ps.

To confirm this possibility and evaluate the plasma conditions driven by different laser/pair flow parameters we have performed detailed 2D PIC simulations with the code OSIRIS [35], capturing the interaction of the relativistic pair plasma from first principles. The interaction is modeled as two millimeter-wide counter-streaming electron-positron flows, with 5 MeV energy ($\gamma_0 = 10$), $T_e = 0.5$ MeV, and flux scaled according to our experimental findings for different laser energies. The simulations use a box size of 400 x 80 $(c/\omega_p)^2$, 10240 x 2048 cells, and 64 particles per cell per species. All simulations used cubic particle shapes, and current and field smoothing with compensation for improved numerical properties.

Figure 3 shows the density and magnetic field structure produced by counter-streaming pair plasmas for two different laser energies, illustrating the shock formation process and the laser-pair flow conditions required for its study. Fig. 3a) corresponds to $E_L = 7$ kJ, $\tau_0 = 10$ ps, N = $10^{13}$, and shows that for these parameters it is possible to reach saturation of the linear stage of the Weibel instability. The magnetic field reaches amplitudes of 0.4 MG, which corresponds to a ratio of magnetic energy density by kinetic energy density of the flows $\sigma = B^2/(16\,\pi\,\gamma_0\,n_0 m_e c^2) \sim 0.4$, illustrating that a significant fraction of the flow energy is converted into magnetic energy, a critical ingredient for the generation of shocks in initially weakly magnetized plasmas. By increasing the laser energy and duration to $E_L = 22$ kJ and $\tau_0 = 25$ ps, which, according to our scaling leads to N = $10^{14}$ per flow, we observe the possibility of reaching shock formation (Fig. 3b), in agreement with Eq. (2). In this case the flows are compressed by a factor of 3, in agreement with the hydrodynamic jump conditions in 2D [36]. The magnetic field associated with the shock reaches amplitudes of mega-gauss and the filaments size is ~0.5 mm, allowing for probing its structure with proton radiography [37].

In summary, we have shown that the positron yield from laser-solid interactions scales with the square of the laser energy, reaching unprecedented high yields. This favorable scaling is due to a combination of increased laser intensity and the recirculation of MeV electrons in the mm-thick target. Our results show that laser-produced pair jets offer ideal experimental conditions to study for the first time the formation of relativistic pair shocks in the laboratory and probe the physics of particle acceleration relevant to high-energy astrophysical environments.

**Acknowledgement:** We acknowledge the support of the Omega EP, Titan and Orion laser facilities for the experiments. Computing support for this work came from ALCC and INCITE awards on Mira (ALCF supported under contract DE-AC02-06CH11357) and from two LLNL Institutional Computing Grand Challenge awards. Additionally, the authors would like to acknowledge the OSIRIS Consortium, consisting of UCLA and IST (Lisbon, Portugal) for the use of the OSIRIS 2.0 framework and the visXD framework. This work was performed under the auspices of the U.S. DOE by LLNL under Contract DE-AC52-07NA27344, and funded by the LDRD (12-ERD-062) program. F. F. acknowledges the financial support from LLNL Lawrence Fellowship.

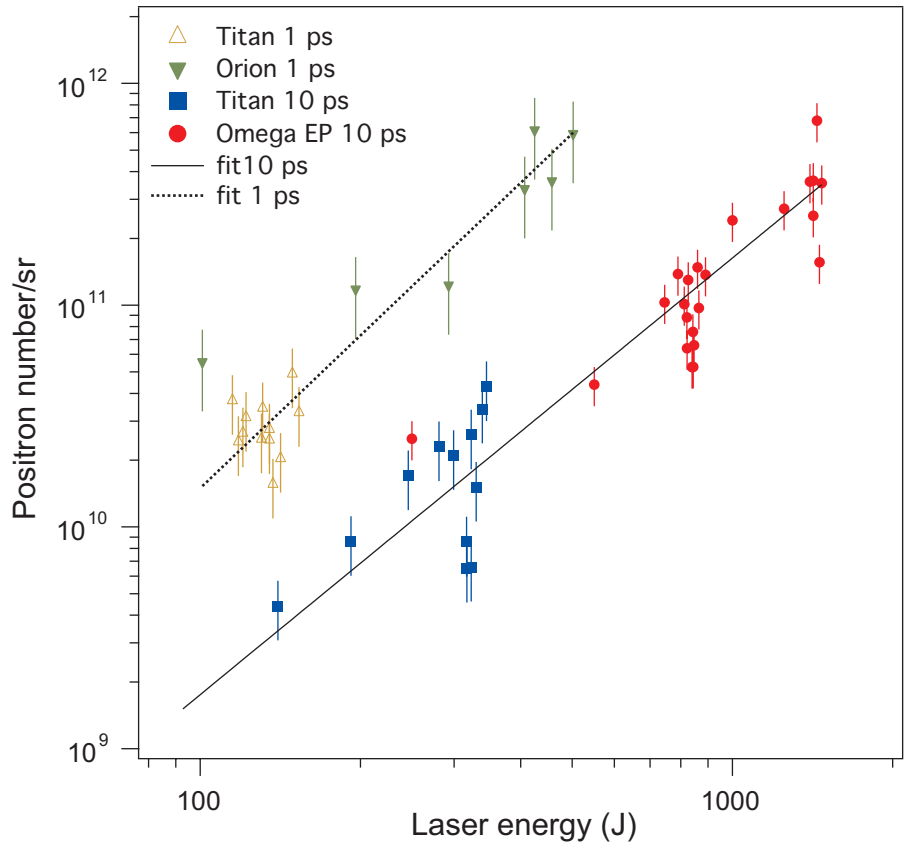

Figure 1: Dependence of the measured positron yield on the laser energy, $E_L$, obtained at three different laser facilities: Omega EP, Orion, and Titan. The upper group is from shots with 1 ps laser pulse: (brown) triangles Titan and (green) diamonds Orion. The lower group is obtained with 10 ps laser pulse: (blue) squares Titan and (red) circles Omega EP.

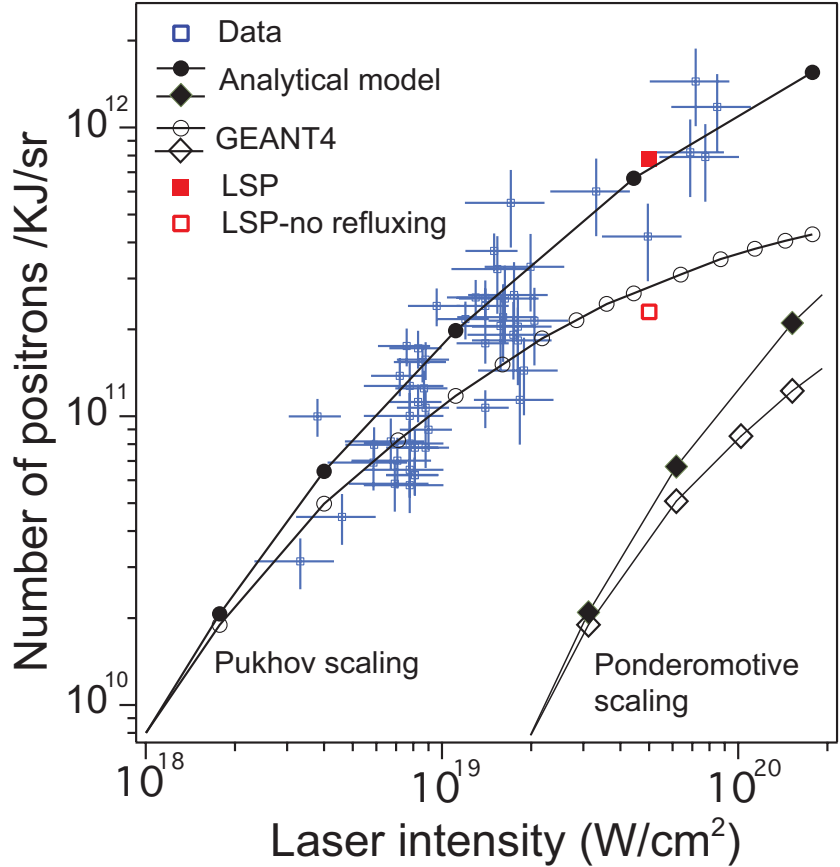

Figure 2: Dependence of the positron yield, per kJ of incident laser energy, on the laser intensity. The experimental data is compared with analytical calculations based on the model of Ref. [20] including refluxing are shown in solid black circles (or diamonds) and with GEANT4 simulations shown in empty black circles (or diamonds). The calculations were made for using two different Te scalings with laser intensity: Pukhov scaling (circles) and Ponderomotive scaling (diamonds). The results from LSP simulations are shown for the cases with refluxing (solid red square) and without refluxing (a hollow square). The data is well fitted when refluxing is included and we use Pukhov scaling for Te. The conversion efficiency from the laser energy to electrons calculated by LSP is about 40%, and the same conversion efficiency is used for both Te scalings in the analytical and GEANT4 calculations.

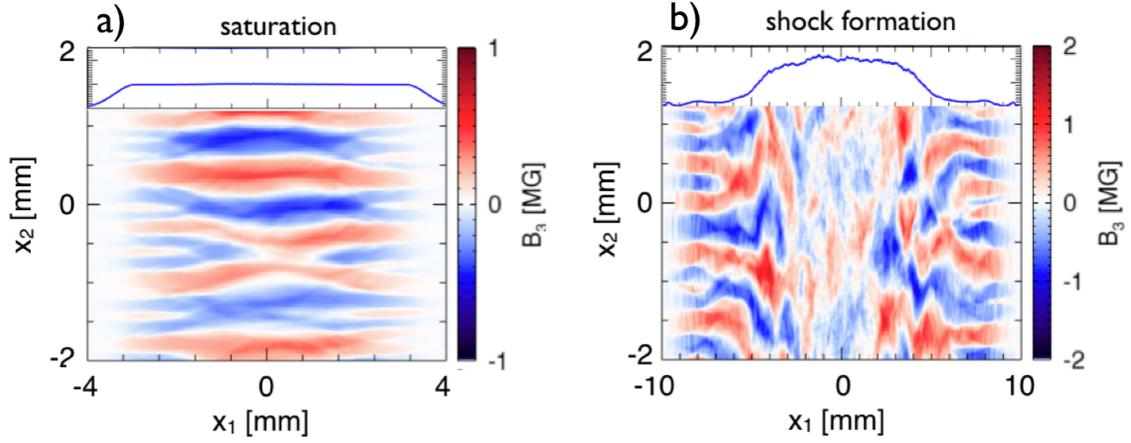

Figure 3. Particle-in-cell simulations of the counter-streaming of relativistic pair plasmas for laser-driven laboratory parameters. Magnetic field structure and transversely averaged density profile (inset) at the end of the interaction are shown for pair plasma flows corresponding to laser parameters of a) $E_L = 7$ kJ, $\tau_0 = 10$ ps, and b) $E_L = 22$ kJ, $\tau_0 = 25$ ps. The results illustrate the possibility of reaching (a) the saturation of the Weibel instability and (b) the formation of a shock with near-future laser systems.